\documentclass[aps,twocolumn,showpacs,footinbib]{revtex4}%
\usepackage[latin9]{inputenc}
\usepackage{amssymb}
\usepackage{amsmath}
\usepackage{amscd}
\usepackage{graphicx}
\usepackage{subfigure}
\usepackage{float}
\begin{document}
\draft
\title{Quantum information processing and multiatom entanglement engineering with a
thermal cavity}
\author{Shi-Biao Zheng\thanks{%
E-mail: sbzheng@pub5.fz.fj.cn}}
\address{Department of Electronic Science and Applied Physics\\
Fuzhou University\\
Fuzhou 350002, P. R. China}

\begin{abstract}
We propose a scheme for realizing two-qubit quantum phase gates with atoms
in a thermal cavity. The photon-number dependent parts in the evolution
operator are canceled with the assistant of a strong classical field. Thus
the scheme is insensitive to the thermal field. In the scheme the detuning
between the atoms and the cavity is equal to the atom-cavity coupling
strength and thus the gates operate at a high speed, which is also important
in view of decoherence. The scheme can be generalized to generate multiatom
entangled states with a thermal cavity.
\end{abstract}

\maketitle
\vskip
0.5cm

\narrowtext

Recently, much attention has been paid to the quantum computers, which are
based on the fundamental principles of quantum mechanics. The new type of
machines can solve some problems much faster than the classical computers,
such as factorizing a large integer [1] and searching for an item from a
disordered system [2]. It has been shown that the building blocks of quantum
computers are two-qubit gates [3]. In cavity QED, schemes have been proposed
for realizing quantum logic gates [3,4]. The ion trap is also a good system
for quantum information processing [5]. Quantum logic gates have been
demonstrated in cavity QED [6,7], ion trap [8], and NMR [9] experiments.

One of the main obstacles for the implementation of quantum information in
microwave cavity QED is the decoherence of the cavity field, while that in
ion traps is the difficulty to control the collective vibrational motion of
the ions. Recently, Schemes have been proposed for realizing quantum
computation in ion traps via virtual vibrational excitations [10,11]. The
schemes does not use the motional mode as the data bus and is insensitive to
the vibrational states. We have proposed a scheme for generation of two-atom
entangled states within a nonresonant microwave cavity [12,13]. The scheme
is insensitive to the thermal field and photon decay although the effective
Hamiltonian involves photon-number dependent Stark shifts, which are
canceled if one atom is initially in the excited state and the other in the
ground state. Following our scheme, an experiment has been reported, in
which two Rydberg atoms crossing a nonresonant cavity are entangled by
coherent energy exchange [14].

In Ref. [12] we also proposed a scheme for the realization of quantum logic
gates and atomic state teleportation. In this case a third auxiliary atomic
state is employed so that the photon-number dependent Stark shifts can not
be canceled. Therefore, the scheme requires that the cavity remain in the
vacuum state throughout the procedure and thus is insensitive to the photon
decay. However, it is sensitive to the thermal field, which builds up during
the operations and relaxes to thermal equilibrium with the increase of the
number of qubits and the number of operations [14]. Thus, the scheme of Ref.
[12] is not feasible when scalability is required. In this paper we propose
an alternative scheme for realizing two-qubit phase gates in cavity QED. The
distinct feature of the present scheme is that the photon-number dependent
parts in the evolution operator are canceled with the assistant of a strong
classical driving field. Due to this feature the scheme is insensitive to
the thermal field. Unlike the previous scheme [12], the present scheme does
not require the detuning between the atoms and the cavity to be much larger
than the atom-cavity coupling strength and thus the operation speed is
greatly improved, which is also important in view of decoherence. Due to
these advantages oue scheme may open promising prospects for complex quantum
information manipulation. The scheme can also be used to generate multiatom
entangled states with a single thermal cavity.

We consider two identical two-level atoms simultaneously interacting with a
single-mode cavity field and driven by a classical field. In the
rotating-wave approximation, the Hamiltonian is (assuming $\hbar =1$) [15]
\begin{eqnarray}\label{1}
&H=\omega _0S_z+\omega _aa^{+}a+\sum_{j=1,2}[\frac g2(a^{+}S_j^{-}+aS_j^{+})&\cr\cr&+%
\frac \Omega 2(S_j^{+}e^{-i\omega t}+S_j^{-}e^{i\omega t})],&
\end{eqnarray}
where S$_j^{+}=\left| e_j\right\rangle \left\langle g_j\right| $, S$%
_j^{-}=\left| g_j\right\rangle \left\langle e_j\right| ,$ $S_z=\frac 12%
\sum_{j=1,2}(\left| e_j\right\rangle \left\langle e_j\right| -\left|
g_j\right\rangle \left\langle g_j\right| ),$ with $\left| e_j\right\rangle $
and $\left| g_j\right\rangle $ (j=1,2) being the excited and ground states
of the jth atom, a$^{+}$ and a are the creation and annihilation operators
for the cavity mode, and g is the atom-cavity coupling strength, $\Omega $
is the Rabi frequency of the classical field, $\omega _0$ is the atomic
transition frequency, $\omega _a$ is the cavity frequency, and $\omega $ is
the frequency of the classical field. Assume that $\omega _0=\omega .$ Then
the interaction Hamiltonian, in the interaction picture, is
\begin{equation}\label{2}
H_i=\sum_{j=1,2}[\frac g2(e^{-i\delta t}a^{+}S_j^{-}+e^{i\delta t}aS_j^{+})+%
\frac \Omega 2(S_j^{+}+S_j^{-})],
\end{equation}
$\delta =\omega _0-\omega _a$. The free Hamiltonian $\omega _0S_z+\omega
_aa^{+}a$ is used for the transformation to the interaction picture. Define
the new atomic basis [11,15]

\begin{equation}\label{3}
\left| +_j\right\rangle =\frac 1{\sqrt{2}}(\left| g_j\right\rangle
+\left| e_j\right\rangle ),\text{ }\left| -_j\right\rangle =\frac
1{\sqrt{2}}(\left| g_j\right\rangle -\left| e_j\right\rangle ).
\end{equation}
Then we can rewrite $H_i$ as
\begin{eqnarray}\label{4}
&H_i=\sum_{j=1,2}\{\frac g2[e^{-i\delta t}a^{+}(\sigma _{z,j}+\frac
12\sigma _j^{+}-\frac 12\sigma _j^{-})&\cr\cr
&+e^{i\delta
t}a(\sigma _{z,j}+\frac 12\sigma _j^{-}-\frac 12\sigma
_j^{+})]+\Omega \sigma _{z,j}\},&
\end{eqnarray}
where $\sigma _{z,j}=\frac 12(\left| +_j\right\rangle \left\langle
+_j\right| -\left| -_j\right\rangle \left\langle -_j\right| ),$
$\sigma _j^{+}=\left| +_j\right\rangle \left\langle -_j\right| $ and
$\sigma _j^{-}=\left| -_j\right\rangle \left\langle +_j\right| .$

The time evolution of this system is decided by Schr\"odinger's equation:
\begin{equation}\label{5}
i\frac{d|\psi (t)\rangle }{dt}=H_i|\psi (t)\rangle .
\end{equation}
Perform the unitary transformation
\begin{equation}\label{6}
|\psi (t)\rangle =e^{-iH_0t}|\psi ^{^{\prime }}(t)\rangle ,
\end{equation}
with
\begin{equation}\label{7}
H_0=\Omega \sum_{j=1,2}\sigma _{z,j}.
\end{equation}
Then we obtain
\begin{equation}\label{8}
i\frac{d|\psi ^{^{\prime }}(t)\rangle }{dt}=H_i^{^{\prime }}|\psi
^{^{\prime }}(t)\rangle ,
\end{equation}
where
\begin{eqnarray}\label{9}
&H_i^{^{\prime }}=\sum_{j=1,2}\frac g2[e^{-i\delta t}a^{+}(\sigma _{z,j}+%
\frac 12\sigma _j^{+}e^{i\Omega t}-\frac 12\sigma _j^{-}e^{-i\Omega
t})&\cr\cr
&+e^{i\delta t}a(\sigma _{z,j}+\frac 12\sigma _j^{-}e^{-i\Omega t}-\frac 12%
\sigma _j^{+}e^{i\Omega t})].&
\end{eqnarray}

Assuming that $\Omega \gg \delta ,g,$ we can neglect the terms oscillating
fast. Then $H_i^{^{\prime }}$ reduces to [15]
\begin{eqnarray}\label{10}
H_i^{^{\prime }} &=&\sum_{j=1,2}\frac g2(e^{-i\delta
t}a^{+}+e^{i\delta
t}a)\sigma _{z,j} \\
\ &=&\frac g2(e^{-i\delta t}a^{+}+e^{i\delta t}a)S_x,  \nonumber
\end{eqnarray}
where

\begin{equation}\label{11}
S_x=\frac 12\sum_{j=1,2}(S_j^{+}+S_j^{-})
\end{equation}
The evolution operator for Hamiltonian H$_i^{^{\prime }}$ can be written in
the form of [16]

\begin{equation}\label{12}
U^{^{\prime }}(t)=e^{-iA(t)S_x^2}e^{-iB(t)S_xa}e^{-iC(t)S_xa^{+}}.
\end{equation}
Using the Schr\"odinger equation

\begin{equation}\label{13}
i\frac{dU^{^{\prime }}(t)}{dt}=H_iU^{^{\prime }}(t),
\end{equation}
we obtain

\begin{equation}\label{14}
B(t)=\int_0^t\frac g2e^{i\delta t^{^{\prime }}}dt^{^{\prime }}=\frac g{%
2i\delta }(e^{i\delta t}-1),
\end{equation}
\begin{equation}
C(t)=\int_0^t\frac g2e^{-i\delta t^{^{\prime }}}dt^{^{\prime }}=\frac g{%
-2i\delta }(e^{-i\delta t}-1),
\end{equation}
\begin{equation}
A(t)=i\int_0^tB(t^{^{\prime }})\frac g2e^{-i\delta t^{^{\prime
}}}dt^{^{\prime }}=\frac{g^2}{4\delta }[t+\frac 1{i\delta }(e^{-i\delta
t}-1)].
\end{equation}

Setting
\begin{equation}
\delta t=2\pi ,
\end{equation}
we have B(t)=C(t)=0. Then we have
\begin{equation}
U^{^{\prime }}(t)=e^{-i\lambda tS_x^2},
\end{equation}
where $\lambda =\frac{g^2}{4\delta }$ The evolution operator of the system
is given by

\begin{equation}\label{19}
U(t)=e^{-iH_0t}U^{^{\prime }}(t)=e^{-i\Omega tS_x-i\lambda tS_x^2}.
\end{equation}
It can be easily shown that [17]
\begin{eqnarray}\label{20}
&U(t)\left| +_1\right\rangle \left| +_2\right\rangle =e^{-i(\Omega
+\lambda )t}\left| +_1\right\rangle \left| +_2\right\rangle ,&\text{
}\cr\cr &U(t)\left| +_1\right\rangle \left| -_2\right\rangle =\left|
+_1\right\rangle \left| -_2\right\rangle ,& \cr\cr &U(t)\left|
-_1\right\rangle \left| +_2\right\rangle =\left| -_1\right\rangle
\left| +_2\right\rangle ,&\text{ } \cr\cr  &U(t)\left|
-_1\right\rangle \left| -_2\right\rangle =e^{-i(\Omega -\lambda
)t}\left| -_1\right\rangle \left| -_2\right\rangle .&
\end{eqnarray}
Choose the interaction time t and Rabi frequency $\Omega $ appropriately so
that
\begin{equation}\label{21}
\lambda t=\pi /2
\end{equation}
and
\begin{equation}
\Omega t=(2k+\frac 12)\pi ,\text{with k being an integer}.
\end{equation}
Then we have
\begin{eqnarray}
U(t)\left| +_1\right\rangle \left| +_2\right\rangle =-\left|
+_1\right\rangle \left| +_2\right\rangle ,\text{ }\cr\cr U(t)\left|
+_1\right\rangle \left| -_2\right\rangle =\left| +_1\right\rangle
\left| -_2\right\rangle , \cr\cr U(t)\left| -_1\right\rangle \left|
+_2\right\rangle =\left| -_1\right\rangle \left| +_2\right\rangle
,\text{ }\cr\cr U(t)\left| -_1\right\rangle \left| -_2\right\rangle
=\left| -_1\right\rangle \left| -_2\right\rangle .
\end{eqnarray}
By this way we obtain a quantum phase gate. Setting $\delta =g$ and gt=2$\pi
$ Eqs. (17) and (21) can be satisfied. We can choose the Rabi frequency $%
\Omega $ appropriately to satisfy (22). In this case the atomic state
evolution operator $U(t)$ is independent of the cavity field state, allowing
it to be in a thermal state. In the previous scheme [12], it is required
that $\delta \gg g$ and g$^2t/(4\delta )=\pi .$ Therefore, the operation
time in the present scheme is much shorter than that in the previous scheme.

We now turn to the problem of generating multiatom entanglement with a
thermal cavity. Multiphoton entanglement has been observed and used to
verify quantum nonlocality [18]. Schemes have been proposed for producing
multiparticle entangled states of hot ions [19,20]. Following the scheme of
Ref. [19] four-particle entanglement has been demonstrated in ion traps
[21]. On the other hand, three-particle entanglement has been demonstrated
in cavity QED [22]. However, in the experiments reported in Refs. [21] and
[22] the fidelity of entangled states needs to be significantly improved in
order to be useful for the test of quantum nonlocality and in quantum
information processing. We have proposed an alternative scheme for the
generation of three-atom entangled states with a single cavity always in the
vacuum state [23], which is insensitive to the photon decay but sensitive to
thermal fields. We here show how we can do so with a single thermal cavity.
We consider N identical two-level atoms simultaneously interacting with a
single-mode cavity field and driven by a strong classical field. In the case
$\delta t=2\pi $ the evolution operator of the system is given by

\begin{equation}
U(t)=e^{-i\Omega tS_x-i\lambda tS_x^2},
\end{equation}
where

\begin{equation}
S_x=\frac 12\sum_{j=1}^N(S_j^{+}+S_j^{-}).
\end{equation}
Assume that the atoms are initially in the state $\left|
g_1g_2...g_N\right\rangle .$ Using the representation of the operator S$_z$,
the atomic state $\left| g_1g_2...g_N\right\rangle $ and $\left|
e_1e_2...e_N\right\rangle $ can be expressed as $\left|
N/2,-N/2\right\rangle $ and $\left| N/2,N/2\right\rangle .$ On the other
hand, such states can be expanded in terms of the eigenstates of S$_x$
[19,24]

\begin{equation}
\left| N/2,-N/2\right\rangle =\sum_{M=-N/2}^{N/2}C_M\left|
N/2,M\right\rangle _x.
\end{equation}
\begin{equation}
\left| N/2,N/2\right\rangle =\sum_{M=-N/2}^{N/2}C_M(-1)^{N/2-M}\left|
N/2,M\right\rangle _x.
\end{equation}
Thus, the evolution of the system is

\begin{equation}
\sum_{M=-N/2}^{N/2}C_Me^{-i(\Omega M+\lambda M^2)t}\left| N/2,M\right\rangle
_x.
\end{equation}
When N is even M is an integer. With the choice $\lambda t=\pi /2$ and $%
\Omega t=2n\pi $ we obtain
\begin{equation}
\begin{array}{c}\label{29}
\left| \psi (t)\right\rangle =\frac 1{\sqrt{2}}\sum_{M=-N/2}^{N/2}C_M[e^{-i%
\pi /4}+e^{i\pi /4}(-1)^M]\left| N/2,M\right\rangle _x \cr\cr =\frac
1{\sqrt{2}}(e^{-i\pi /4}\left| N/2,-N/2\right\rangle +e^{i\pi
/4}(-1)^{N/2}\left| N/2,N/2\right\rangle ).\\
\end{array}
\end{equation}
On the other hand, for the case that N is odd we can set $M=M^{^{\prime
}}+1/2,$ with $M^{^{\prime }}$ being an integer. With the choice $\lambda
t=\pi /2$ and $\Omega t=(4n+\frac 32)\pi $ we obtain
\begin{equation}
\begin{array}{c}\label{30}
\left| \psi (t)\right\rangle =\frac 1{\sqrt{2}}e^{i\frac 78\pi
}\sum_{M=-N/2}^{N/2}C_M[e^{-i\pi /4}+e^{i\pi /4}(-1)^{M^{^{\prime
}}}]\cr\cr\left| N/2,M\right\rangle _x \cr\cr =\frac
1{\sqrt{2}}e^{i\frac 78\pi }[e^{-i\pi /4}\left|
N/2,-N/2\right\rangle \cr\cr +e^{i\pi /4}(-1)^{(1+N)/2}\left|
N/2,N/2\right\rangle ].
\end{array}
\end{equation}
By this way we obtain multiatom Greenberger-Horne-Zeilinger states [25],
which is useful for test of quantum mechanics.

It should be noted that though the atomic state evolution is independent of
the thermal photons of cavity field after a period t decided by Eq. (17),
the atomic system is entangled with the cavity mode during the atom-cavity
interaction. Therefore, it is required that the cavity decay is negligible
during the interaction. In the experiment reported in Ref. [14] the
atom-cavity coupling strength is about $g=2\pi \times 50kHz$ and thus the
interaction time is on the order of $\pi /g\simeq 10^{-5}s$. The photon
decay time is T$_c\simeq 10^{-3}s,$ much longer than the interaction time.
After the interaction, the atoms are disentangled with the cavity field and
then the cavity decay will not affect the gate operation.

In order to derive Eq. (10) we have assumed that $\Omega \gg \delta $, $g$
and thus discarded the terms
\begin{eqnarray}\label{31}
\Delta H_i^{^{\prime }}=\sum_{j=1,2}\frac g2[e^{-i\delta t}a^{+}(\frac 12%
\sigma _j^{+}e^{i\Omega t}-\frac 12\sigma _j^{-}e^{-i\Omega
t})\cr\cr +e^{i\delta t}a(\frac 12\sigma _j^{-}e^{-i\Omega t}-\frac
12\sigma _j^{+}e^{i\Omega t})].
\end{eqnarray}
These terms induce Stark shifts on the states $\left|
+_j\right\rangle $ and $\left| -_j\right\rangle .$ The Stark shifts
for $\left| +_j\right\rangle $
and $\left| -_j\right\rangle $ are on the order of $g^2/(10\Omega )$ and $%
-g^2/(10\Omega )$, respectively. We here take an example to estimate the
error introduced by the Stark shifts. For the generation of two-atom
maximally entangled state the fidelity is decreased by $\Delta F_1\simeq 1-%
\frac 14\{1+cos[g^2t/(5\Omega )]\}^2$. Setting $\Omega =5\delta ,$ We have $%
\Delta F_1\simeq 0.03$.

The present scheme requires that two atoms be simultaneously sent through a
cavity, otherwise there will be an error. Assume that during the generation
of two-atom maximally entangled state one atom enters the cavity 0.01$t$
sooner than another atom. In this case the fidelity is decreased by $\Delta
F_2\simeq sin^2(0.01\Omega t/2)+sin^2(0.91\lambda t)\simeq 0.02$. Suppose
the fluctuation of the Rabi frequency $\Omega $ is $\Delta \Omega
=0.01\Omega $. This fluctuation also decreases the fidelity by $\Delta
F_3\simeq sin^2(0.01\Omega t/2)\simeq 0.02$.

In conclusion, we have proposed a scheme for realizing two-qubit phase gates
in cavity QED. The scheme is insensitive to the thermal field and works in a
fast way, which is of importance from the experimental point of view. Our
scheme may offer a viable way to built a scalable quantum computer. Our
scheme can also be used to produce multiatom entangled states with a single
thermal cavity.

This work was supported by Fok Ying Tung Education Foundation, the National
Fundamental Research Program Under Grant No. 2001CB309300, the National
Natural Science Foundation of China under Grant No. 60008003, and the
Outstanding Young Scientist Award from the National Natural Science
Foundation of China.


\begin{references}
\bibitem{1}  P. W. Shor, in Proceedings of the 35th Annual Symposium on
Foundations of Computer Science (IEEE Computer Society, Los Alamitos, CA,
1994), P.116.

\bibitem{2}  L. K. Grover, Phys. Rev. Lett. 79 (1997) 325; {\sl ibid.}79
(1997) 4709.

\bibitem{3}  T. Sleator and H. Weinfurter, Phys. Rev. Lett. 74, 4087 (1995).

\bibitem{4}  A. Barenco et al., Phys. Rev. Lett. 74, 4083 (1995); P. Domokos
et al., Phys. Rev. A52, 3554 (1995); T. Pellizzari, S. A. Gardiner, J. I.
Cirac, and P. Zoller, Phys. Rev. Lett. 75, 3788 (1995).

\bibitem{5}  J. I. Cirac and P. Zoller, Phys. Rev. Lett. 74, 4091 (1995); C.
Monroe et al., Phys. Rev. A 55, R2489 (1997).

\bibitem{6}  Q. A. Turchette et al., Phys. Rev. Lett. 75, 4710 (1995).

\bibitem{7}  A. Rauschenbeutel et al., Phys. Rev. Lett. 83, 5166 (1999).

\bibitem{8}  C. Monroe et al., Phys. Rev. Lett. 75, 4714 (1995).

\bibitem{9}  N.A. Gershenfeld and I. L. Chuang, Science 275, 350 (1997); D.
G. Cory et al., Proc. Natl. Acad. Sci. U. S. A. 94, 1634 (1997); J.A. Jones
et al., Nature (London) 393, 344 (1998).

\bibitem{10}  A. S$\phi $rensen and K. M$\phi $lmer, Phys. Rev. Lett. 82, 1971
(1999).

\bibitem{11}  D. Jonathan and M. B. Plenio, Phys. Rev. Lett. 87, 127901 (2001).

\bibitem{12}  S. B. Zheng and G. C. Guo, Phys. Rev. Lett. 85, 2392 (2000).

\bibitem{13}  S. B. Zheng and G. C. Guo, Phys. Rev. A 63, 044302 (2001).

\bibitem{14}  S. Osnaghi et al., Phys. Rev. Lett. 87, 037902 (2001).

\bibitem{15}  E. Solano et al., quant-ph/0202071.

\bibitem{16}  A. S$\phi $rensen and K. M$\phi $lmer, Phys. Rev. A 62, 022311
(2000).

\bibitem{17}  X. Zou et al., quant-ph/0204171.

\bibitem{18}  J-W. Pan et al., {\sl Nature }{\bf 403}, 515 (2000).

\bibitem{19}  K. M$\phi $lmer and A. S$\phi $rensen, Phys. Rev. Lett. 82, 1835
(1999).

\bibitem{20}  S. B. Zheng, Phys. Rev. A 65, 051804 (R) (2002).

\bibitem{21}  C. A. Sackett et al., {\sl Nature} {\bf 404}, 256 (2000).

\bibitem{22}  A. Rauschenbeutel et al., Science 288, 2024 (2000).

\bibitem{23}  S. B. Zheng, Phys. Rev. Lett. 87, 230404 (2001).

\bibitem{24}  D. M. Brink and G. R. Satchler, Angular Momentum (Clarendon
Press, Oxford, 1975).

\bibitem{25}  D. M. Greenberger et al., in {\sl Bell,s Theorem, Quantum
Theory, and Conceptions of the Universe, }edited by M.Kafatos (Kluwer,
Dordrecht, 1989); D. M. Greenberger et al., Am. J. Phys. 58, 1131 (1990).
\end{references}
\end{document}